\newcommand{\br}{{\bf r}}
\documentstyle[aps,twocolumn,epsfig,graphics]{revtex}
\begin{document}
\twocolumn[\hsize\textwidth\columnwidth\hsize
\csname@twocolumnfalse%
\endcsname
\draft
\title {Nonlinear excitations in arrays of Bose-Einstein condensates}
\author{F.Kh. Abdullaev$^{1\dag}$, B.B. Baizakov$^{1\ddag}$,
        S.A. Darmanyan$^2$, V.V. Konotop$^3$, and M. Salerno$^4$}
\address{$^1$Physical-Technical Institute, Uzbek Academy of Sciences,
         2-b, Mavlyanov str., 700084, Tashkent, Uzbekistan \\
         $^2$Institute of Spectroscopy, Russian Academy of Sciences,
         Moscow reg., Troizk,  142092,Russia\\
         $^3$Departmento de F\'{\i}sica and
         Centro de F\'{\i}sica da Mat\'eria
         Condensada, Universidade de Lisboa, \\
         Complexo Interdisciplinar, Av. Prof. Gama Pinto 2, Lisboa 1649-003,
         Portugal \\
         $^4$Dipartimento di Fisica "E.R. Caianiello"
         and Istituto Nazionale di Fisica della Materia (INFM),\\
         Universit\'a di Salerno, via S. Allende,
         I-84081 Baronissi (SA), Italy }
\date{\today}
\maketitle

\begin{abstract}
The dynamics of localized excitations in array of Bose-Einstein
condensates is investigated in the framework of the nonlinear lattice
theory. The existence of temporarily stable ground states
displaying an atomic population distributions localized on very
few lattice sites (intrinsic localized modes), as well as, of
atomic population distributions involving  many lattice sites
(envelope solitons), is studied both numerically and analytically.
The origin and properties of these modes are shown to be
inherently connected with the interplay between macroscopic
quantum tunnelling and nonlinearity induced self-trapping of atoms
in coupled BECs. The phenomenon of Bloch oscillations of these
excitations is studied both for zero and non zero backgrounds. We
find that in a definite range of parameters, homogeneous
distributions can become modulationally unstable. We also show
that bright solitons  and excitations of shock wave type can exist
in BEC arrays even in the case of positive scattering length.
Finally, we argue that BEC array with negative scattering length
in presence of linear potentials can display collapse.
\end{abstract}
\vskip 0.5cm
]

\section{Introduction}

The realization of Bose-Einstein condensates (BEC) and the
observation of quantum interference phenomena between two coupled
condensates has opened a new fascinating field in physics, with
the perspective of getting a better understanding of the
complicate behavior of quantum many body systems and with the hope
of realizing novel concrete applications of quantum mechanics such
as atom interferometers and atom lasers. The existence of periodic
localized oscillations in the relative atomic population of one of
two coupled condensates (Bloch oscillations), was first predicted
in Refs.~\cite{Shen,Melburn,Zapata}. The close analogy existing
with the Josephson effect  was also emphasized, the common origin
of these phenomena being the temporal interference of two
macroscopic quantum states, leading to current oscillations in
Josephson junctions \cite{Likharev}, and to oscillations of the
atomic population in coupled BECs \cite{Wil}.

Although the  generalization of these results to the case of three
coupled BECs was considered \cite{Zhang}, few theoretical
investigations exist on interference phenomena in  arrays of
coupled BEC's \cite{dalfovo}. In a recent paper
\cite{trombettoni}, the problem of Bloch oscillations of bright
solitons was investigated in terms of a tight-binding model for
BEC arrays with positive scattering length. It is a fortunate
situation that the equations arising in this case, formally
coincide with those studied in the theory of nonlinear lattices
\cite{scott}. One can indeed transpose the knowledge gained in
these fields, to the field of BEC arrays. Thus, for example, one
can expect that the types of elementary excitations which can
arise in BEC arrays, as well as their dynamical behavior, strongly
depend on the ratio between the coupling of adjacent condensates
and the nonlinearity induced by the inter atomic interactions.
Keeping the same terminology used in nonlinear lattices theory,
one can say that for weak coupling and strong nonlinearity, {\em
intrinsic localized modes} (ILM), i.e. matter excitations
localized on few lattice sites,  should exist. These modes could
arise in experiments like the ones reported in Ref.~\cite{Andr} in
which millions of atoms were trapped in an almost one-dimensional
1D geometry ("cigar" shape geometry), with a small tunnelling term
between condensates.

On the other hand, when the coupling constant is comparable with
the self-trapping interactions, small amplitude excitations of
large size (in comparison with the lattice constant), should
arise. These excitations can be identified as lattice {\em
envelope solitons} (ES) and could be observable in macroscopic
quantum interference experiments like the ones reported in
Ref.~\cite{Kasevich} in which, vertical arrays of "pancake"-like
BECs, each containing thousands of $^{87}$Rb atoms, were coupled
through the gravitational field. In this case the interference
phenomenon shows up as Bloch oscillations of the ES describing the
atomic population along the array (ES dynamical localization).
Bloch oscillations of ES type are also possible in horizontal
optical lattices induced by two counter-propagating laser beams
with a frequency detuning varying linearly in time \cite{Niu}
(this produces a linear potential on the BEC similar to the
gravitational field of the vertical traps). In the following we
shall call both types of excitations (i.e. ILM or ES) as {\em
discrete matter solitons} (DMS) whenever it will be clear from the
context to which type we refer \cite{commento1}.

The aim of this paper is twofold. From one side, we wish to make a
bridge from the theory of nonlinear lattices to the field of BEC
arrays. This will allow to get a better understanding of the types
of excitations which can arise in 1D coupled BECs as the physical
parameters are varied. From the other side, we wish to expand both
theoretically and numerically, on the phenomenon of Bloch
oscillations in the case of 1D BEC arrays in presence of uniform
backgrounds (with the term ``background" here we mean the presence
of a uniform distribution of atomic population along the array).
In particular, we discuss:

\begin{itemize}
\item [{i)}] the existence of ILM in 1D BEC arrays with
positive scattering length;

\item [{ii)}] the existence of ES in BEC arrays with positive scattering
lengths;

\item [{iii)}]Bloch oscillations of soliton-like excitations on top of
uniform backgrounds in periodic BECs with positive scattering
lengths;

\item [{iv)}]the dynamics of  BEC arrays with
negative scattering length on zero backgrounds.
\end{itemize}

As to point i) we remark that in contrast with the solitons and
breathers discussed of Ref.~\cite{trombettoni} which are extended
on many lattice sites \cite{commento2}, the bright ILM discussed
here are highly discrete, i.e. localized on few lattice sites.
Since ILM have not been  observed in real BEC arrays, this is an
implicit call for experimental work to be done in this direction
(for experimental evidence of ILM existence in other physical
contexts see \cite{Den}). Point ii) generalizes the results of
\cite{trombettoni} to the case of non-zero backgrounds, while in
iii) we show that modulational instability can arise as the slope
of the linear potential (gravitational field for vertical arrays)
is increased. Moreover, we find that for certain parameter values,
shock wave excitations can develop. In point vi) we discuss Bloch
oscillations of bright solitons both analytically and numerically.
All these results  will have a close analogy with those derived in
the theory of nonlinear lattices on intrinsic localized modes
\cite{Siv,Dar,Kiv,Fla}, envelope solitons \cite{CKV,KS1,KonTak},
shock waves \cite{KS1,SMK}, and Bloch oscillations
\cite{Scharf,BLR}.

The paper is organized as follows. In Section II the model of a 1D
array of BEC is derived. In Section III we discuss the existence
of intrinsic localized BEC solitons, both of even and odd
symmetry. In Section IV we use a multi-scale expansion to discuss
the stability and dynamics of small amplitude excitations of BEC
arrays with positive scattering length against a uniform
background and in presence of a linear potential modelling the
gravitational field. The linear stability analysis of the
background for BEC arrays is investigated, and the region in
parameter space where modulational instabilities develop is
provided. We show that for particular parameters, shock wave
propagation can also develop. In section V the main characteristic
of Bloch oscillations for arrays with negative scattering length,
are discussed. Finally, in the last section VI the main results of
the paper are summarized.

\section{The model}

As is well known, in the mean field approximation the wave
function of a  BEC in  a trap potential $V({\bf r})$ satisfies
the time-dependent Gross-Pitaevskii equation (GPE)
\begin{equation} \label{GPE}
  i\hbar \frac{\partial \Psi({\bf r},t)}{\partial t} =
  \bigl[ -\frac{\hbar^2}{2m}\Delta + V({\bf r}) +
  g_0 |\Psi({\bf r},t)|^2 \, \bigr] \Psi({\bf r},t),
\end{equation}
where $g_0 = 4\pi \hbar^2 a_s /m$, $m$ is the atomic mass, $a_s$
is the $s$-wave scattering length which can be either {\it
positive} (the case of $^{87}$Rb atoms, with repulsive
interactions, and with $a_s = 5.5$ nm) or {\it negative} (the case
of $^{7}$Li atoms, with attractive interactions and $a_s =- 1.45$
nm). The trap potential is assumed to be periodic along the
$z$-direction
\begin{equation} \label{potgen}
  V(\br)=V(\br+\lambda\hat{k}),
\end{equation}
where $\lambda$ is the spatial period of the potential and
$\hat{k}$ is a unitary vector in the $z$-direction. A typical
model potential is
\begin{equation} \label{potential}
  V(\br)= mgz + V_0(x,y)\sin^2(kz),
\end{equation}
with experimental parameters $\lambda = 2\pi/k \sim 850\ nm$, $V_0
\sim 2.1 \hbar^2 k^2/2m$, and with an atomic population in each
wells of $N_0 \approx 10^3$ atoms \cite{Kasevich}.

To make analytical studies it is convenient to look for solutions
of Eq.~(\ref{GPE}) of the form
\begin{equation} \label{sup}
  \Psi({\bf r},t) = \sum_{{n=-\mathcal N}/2}^{{\mathcal N}/2}
  \psi_{n}(t)\Phi_{n}({\bf r}),
\end{equation}
where the summation is performed over the number, ${\mathcal N}$,
of minima of the trap potential which is assumed to be even. The
functions $\Phi_n(\br)\equiv\Phi (\br-\br_n)$ are assumed to be
strongly localized around the site $n$ of the potential and
normalized to the mean number of atoms in the n-th well
$N_0=N/{\cal N}_w$,
\begin{equation} \label{normalization1}
  \int\bar{\Phi}_n(\br)\Phi_{n}(\br)\,d\br=N_o,
\end{equation}
where $N$ is the total number of atoms and ${\cal N}_w={\cal N}+1$
the total number of potential wells (hereafter the overbar will
denote complex conjugation). This implies that the hopping
integral
\begin{equation} \label{hopping}
  |J_{n,n+1}|=\Bigg|\int\bar{\Phi}_n(\br)\Phi_{n+1}(\br)\,d\br\Bigg| =
\ll N,
\end{equation}
can be neglected, i.e. we assume  $J_{n,n+1}\approx 0$ for all
$n$. This is analogous to the well known tight binding
approximation of solid state physics \cite{Kittel}, valid for
weakly overlapped condensates. Note that from
(\ref{normalization1}) and (\ref{hopping}) the following
normalization condition for the functions $\psi_n(t)$ is obtained
\begin{equation} \label{normalization2}
  \sum_{n=-{\mathcal N}/2}^{{\mathcal N}/2}|\psi_n(t)|^2=1.
\end{equation}
Substituting (\ref{sup}) into (\ref{GPE}), multiplying by
$\bar{\Phi}_{n}({\bf r})$ and integrating over $\br$, we readily
obtain the system of coupled equations
\begin{eqnarray} \label{DNLS}
  i\hbar\dot\psi_{n} &=& E_n \psi_n + U_n |\psi_n|^2 \psi_n -
  \nonumber \\
  & &  K_{n,n-1} \psi_{n-1} - K_{n,n+1}\psi_{n+1}+\gamma_{0} n \psi_n ,
\end{eqnarray}
where the overdot  means time derivative, and $\gamma_0 = mga$
with $a = \lambda/2$ the distance between adjacent wells. In
Eq.~(\ref{DNLS}) we have denoted with
\begin{equation} \label{E}
  E_n = \int \bigl[ \frac{\hbar^2}{2m}|\nabla \Phi_n|^2 + |\Phi_n|^2
          V({\bf r}) \bigr] \, d{\bf r}
\end{equation}
the zero-point energy for the well $n$, with
\begin{equation} \label{U}
U_n = \frac{g_0}{N} \int |\Phi_{n}|^4 \, d{\bf r}
\end{equation}
the mean-field self-interaction energy, and with
\begin{equation} \label{K}
K_{n, n\pm 1}= -\int \bigl[ \frac{\hbar^2}{2m}
(\nabla\Phi_n \nabla \Phi_{n\pm 1})+(\Phi_n V\Phi_{n\pm1})\bigr] \,
d{\bf r}
\end{equation}
the sum of the kinetic energy and the off-diagonal matrix element
of the trap potentials  between sites $n$ and $n+1$ (i.e. the
coupling constant between neighboring BECs). In the following we
shall consider the case of equal constants $E_n = E,\ U_n=U, \
K_{n,n\pm 1} = K$. Introducing the new variables
\begin{equation} \label{renorm}
  t \rightarrow \frac{K}{\hbar}t, \qquad \psi_n
  \rightarrow \bigl( \frac{2|U|}{K}\bigr)^{1/2}\bar{\psi_n}
  e^{i(\frac{E}{\hbar K}- 2)t}.$$
\end{equation}
we can rewrite Eq.~(\ref{DNLS}) in dimensionless form as
\begin{equation}\label{DNLS1}
  i\dot \psi_n + \psi_{n+1} + \psi_{n-1}  - 2\psi_n + 2\sigma |\psi_n
  |^2 \psi_n  + \gamma n \psi_n= 0,
\end{equation}
with $\sigma = -\mbox{sign}(a_s)$ and $\gamma = \gamma_0 /K$.
Equation (\ref{DNLS1}) coincides with the well known discrete
nonlinear Schr\"odinger equation \cite{scott}, extensively
investigated in the theory of nonlinear lattices. In the present
context, however, because of the normalization condition
(\ref{normalization2}), the boundary conditions for
Eq.~(\ref{DNLS1}) cannot be choosen arbitrarily. In this paper we
consider either periodic boundary condition
\begin{equation} \label{period}
  \psi(n)=\psi(n+{\cal N}+1)
\end{equation}
in the case of finite lattices, or zero
\begin{equation} \label{zero}
  \lim_{|n|\to \infty}\psi(n)=0
\end{equation}
or "finite density"
\begin{equation} \label{finden}
  \lim_{n\to\pm \infty}\psi(n)=\rho e^{\pm \vartheta}
\end{equation}
boundary conditions, in the case of infinite lattices (here $\rho$
and $\vartheta$ are arbitrary constants). Eq.~(\ref{DNLS1}) will
then have also another integral of motion of the form
\begin{equation} \label{H1}
  H = \sum_{n}\bar{\psi_n}\psi_{n-1}-|\psi_n|^4.
\end{equation}

The mean field tight-binding model in Eq.~(\ref{DNLS1}) will be
used in the following to study the dynamical properties of both
ILM and ES excitations in 1D BEC arrays.

\section{Existence of Intrinsic localized modes in BEC arrays}

As is well known, in the case of continuous 1D BECs, soliton
ground state solutions can exist. These were theoretically
predicted in \cite{Reinhardt} and experimentally observed in the
case of positive scattering length (dark solitons)
\cite{Denschlag}. Bright solitons can exist in BECs only when the
effects of focusing nonlinearity balance the ones of effective
dispersion, this being possible for attractive inter-particle
interactions or negative scattering lengths, only.  Since bright
solitons are very important for developing BEC applications
\cite{Pot}, it is of interest to investigate their existence also
in BEC arrays. In this case, as we will see in the following, new
possibilities can arise. In Ref.~\cite{trombettoni} bright
solitons of the ES type were numerically investigated in 1D BEC
arrays with positive scattering lengths. Here we shall consider
the case of bright solitons of ILM type in BEC arrays with both
positive and negative scattering lengths. We fix $\gamma =0$ ,
$\sigma=1$, in Eq.~(\ref{DNLS1}), i.e. we consider BECs in
horizontal traps with negative scattering lengths. ILM are then
expected to exist when the coupling constant (which is responsible
in the linear case of the spreading of the BEC wavefunction) is
small in comparison with the nonlinear self-trapping interaction.
From the normalization condition (\ref{normalization1}), and from
definitions (\ref{U}), (\ref{K}), one can estimate $|U|/K=O(N)$,
i.e. a stronger nonlinearity corresponds to a larger number of
particles. For the sake of analytical developments is convenient
to introduce the small parameter
\begin{equation} \label{klim}
  \kappa=\sqrt{\frac{K}{|U|}}=O(N^{-1/2})\ll 1
\end{equation}
representing the ratio between tunnel coupling and nonlinearity,
and recast Eq.~(\ref{DNLS1}) it in the form:
\begin{equation} \label{DNLS2}
  i\dot\varphi _{n}+\kappa (\varphi _{n+1}+\varphi _{n-1})+\left|
  \varphi_{n} \right|^{2}\varphi_{n} = 0,
\end{equation}
with $\varphi_{n}=\sqrt{\kappa}\psi_{n}e^{2it}$ (the overdot here
denotes the derivative with respect to $ \tau =t/\kappa$).

In the following we discus ILM solutions of Eq.~(\ref{DNLS2}) with
different symmetry properties.

\subsection{Symmetric ILM centered on a site}

\noindent Symmetric ILM solutions centered on a site (say $n=0$)
\begin{equation} \label{supposition1}
  \varphi_{-n}(\tau)=\varphi_{n}(\tau), \qquad \qquad
  |\varphi_0(\tau)|=1,
\end{equation}
can be searched in the form
\begin{eqnarray} \label{expanphi}
  \varphi_n=e^{i\omega \tau}\sum_{j=n}^{\infty}\varphi_{nj}\kappa^j
  \,, \qquad  \omega=\sum_{j=0}^{\infty}\omega_j \kappa^j,
\end{eqnarray}
with the frequency of the local oscillation $\omega$ equal for all
sites. Substituting this expansion into Eq.~(\ref{DNLS2}) and
performing straightforward algebra, we obtain
\begin{equation} \label{omegaeven}
  \omega=1+2\kappa^2-2\kappa^6+2\kappa^8+O(\kappa^{10}),
\end{equation}
with the site amplitude  $\varphi_n$ satisfying the lattice
equations
\begin{eqnarray}
  \varphi_1&=&[\kappa-\kappa^5+\kappa^7+O(\kappa^9)]e^{i\omega \tau},
  \nonumber \\
  \varphi_2&=&[\kappa^2-\kappa^4+\kappa^8+O(\kappa^{10})]e^{i\omega \tau},
  \nonumber \\
  \varphi_3&=&[\kappa^3-2\kappa^5+\kappa^7+O(\kappa^9)]e^{i\omega \tau},
  \nonumber \\
  \varphi_4&=&[\kappa^4-3\kappa^6+O(\kappa^8)]e^{i\omega \tau},
  \nonumber \\
  \varphi_5&=&[\kappa^5+O(\kappa^7)]e^{i\omega \tau}, \nonumber  \\
  \nonumber
  &\vdots&
\end{eqnarray}
from which we see that $\left| \varphi_{n}\right| \gg \left|
\varphi_{n+1}\right|$ provided $\kappa\ll 1$.

It is remarkable  that the decay of the amplitude $\varphi_n$ is
exponential as one move away from the $n=0$ site, and in the
leading order can be approximated by
\begin{equation} \label{slm1}
  \varphi_{n}\approx\exp (i\omega \tau -\eta \left| n\right|),
\end{equation}
with $\eta$ given by $\exp(-\eta )= \kappa << 1$. Note that in
accordance with our scaling, the limit $\kappa\to 0$ can be viewed
as equivalent to the strongly nonlinear limit $N\to\infty$ of the
original physical system.  Moreover, from Eq.~(\ref{slm1}) it
follows that a bright onsite centered ILM can have for
$\kappa=0.1$ about $91$ \% of all the atoms concentrated on its
central site. For negative scattering lengths this implies that
ILM can be stable only if $N<N_{cr}$, where $N_{cr}$ is the
critical threshold at which collapse phenomena occur (the
effective coupling constant cannot be made smaller than a critical
value, this being a feature of BEC arrays with respect to other
nonlinear lattices).

To check these results we have performed numerical integrations of
Eq.~(\ref{DNLS2}) using as initial condition Eq.~(\ref{slm1}) with
$\tau=0$. We used a DOPRI8 integration routine \cite{Hairer},
based on a Runge-Kutta scheme of 7th-8th order with automatic
stepsize control, so to combine speed with high precision (the
relative errors was  from $10^{-7}$ up to $10^{-13}$). The
integration domain was taken large enough ($800$ sites) to avoid
the influence of boundary conditions. The results are presented in
Figs.\ref{figure1}a,b for the case, respectively, of weak and
strong coupling among neighboring BECs in the array. We see that
in the case of weak coupling the initial atomic population remains
stable for arbitrary long times (Fig. 1a), while for strong
couplings, the initial distribution of atoms spreads out over the
whole system (Fig. 1b).
\begin{figure}[h]
\caption{Distribution of the normalized atomic population
according to numerical solution of Eq.~(\ref{DNLS2}) for the case
of weak coupling, $\kappa=0.1$, (a) and strong coupling,
$\kappa=0.8$ (b): this distribution corresponds to $\tau=10$. The
dashed line shows the ILM envelope given by Eq.~(\ref{slm1}).}
\label{figure1}
\end{figure}
\subsection{Anti-symmetric ILM centered on a site}

\noindent ILM of anti-symmetric type
\begin{eqnarray} \label{symodd}
  \varphi_{0} = 0,\qquad \varphi_1=1, \qquad \varphi_{-n}=-\varphi_{n},
\end{eqnarray}
can be searched in the form ($n>1$)
\begin{eqnarray} \label{expanphi1}
  \varphi_n=e^{i\omega =
  \tau}\sum_{j=n}^{\infty}\varphi_{nj}\kappa^{j-1} \,, \qquad
  \omega=\sum_{j=0}^{\infty}\omega_j \kappa^j.
\end{eqnarray}
Direct substitution into Eq.~(\ref{DNLS2}) gives
\begin{equation} \label{omegaodd}
  \omega=1+\kappa^2+\kappa^4+2\kappa^6+6\kappa^8+O(\kappa^9),
\end{equation}
\begin{eqnarray}
  \varphi_2&=&[\kappa+\kappa^3+2\kappa^5+6\kappa^7+
  O(\kappa^9)]e^{i\omega \tau}, \nonumber \\
  \varphi_3&=&[\kappa^2+\kappa^4+2\kappa^6+5\kappa^8+
  O(\kappa^{10})]e^{i\omega \tau}, \nonumber \\
  \varphi_4&=&[\kappa^3+\kappa^5+\kappa^7+
  O(\kappa^9)]e^{i\omega \tau}, \nonumber \\
  \varphi_5&=&[\kappa^4+\kappa^6+O(\kappa^8)]e^{i\omega \tau}, \nonumber
  \\ \nonumber
  \vdots
\end{eqnarray}

In analogy with the symmetric case, one can ensure that to
leading order the shape of the ILM is well described by the
function
\begin{equation}
  \varphi_{n}\approx \exp (i\omega \tau -\eta \left| n-1\right|),\qquad
  n \geq 1.  \label{slm2}
\end{equation}
Direct numerical simulations of Eq.~(\ref{slm1}) with initial
conditions given by Eq.~(\ref{slm2})), showed a behavior  of the
atomic population distribution as a function of the coupling
constant, similar to the one reported in Fig.~\ref{figure1}.

\bigskip

Other types of bright symmetric and anti-symmetric ILMs,  such as
the ones which are centered between two sites, can also be
constructed. A more detailed analysis of these excitations shows
that the stability of ILMs depends on their symmetry property, as
well as, on their localization extension \cite{Dar,Fla,LKS}. Thus,
symmetric on-site centered ILMs are stable but symmetric centered
between sites ILMs are unstable, while antisymmetric centered
between sites ILM are stable if $\kappa<\kappa_{cr}\approx 0.12,$
and unstable if \ $\kappa>\kappa_{cr}.$

For positive scattering lengths, different families of ILMs (as
well as discrete fronts) representing dark ILM (i.e. a constant
atomic population along the array except for few sites in which
the population is decreased), can also be found
\cite{Dar,Kiv,KonTak}.

It is important to note  that, due to the invariance of
Eq.~(\ref{DNLS2}) under the transformation $\varphi_{n}\rightarrow
(-1)^{n}\varphi _{n}$, $\kappa \rightarrow -\kappa$,
we have that the above bright (dark) ILMs, multiplied by
a factor $(-1)^{n}$, are also solutions of Eq.~(\ref{DNLS2}) for
$ \ a_s>0$ $(\ a_s<0)$. This shows Eq.~(\ref{DNLS2}) possesses both
bright and dark ILM solutions, for any sign of $a_s.$

\section{BEC arrays with positive scattering lengths}

From the experiment in Ref.~\cite{Kasevich} on vertical BEC arrays
and from numerical simulations, it is known that the phenomenon of
Bloch oscillations of ES can occur. In this section we shall use
methods of nonlinear lattices theory \cite{Scharf,BLR,KCV,KonTMF}
to investigate Bloch oscillations of ES in BEC arrays with
positive scattering lengths. For bright ES on zero backgrounds
this was done in Ref.~\cite{trombettoni}, so we concentrate here
only  on the case of  ES on top of non zero backgrounds. To this
end it is convenient to introduce the background amplitude $\rho$
directly into Eq.~(\ref{DNLS1}) with $\sigma=-1$, this giving
\begin{eqnarray} \label{e1}
  i\dot{\varphi}_n + \varphi_{n+1}+\varphi_{n-1}-2\varphi_n &+& \\
  \nonumber
  2(\rho^2-|\varphi_n|^2)\varphi_n &+& \gamma n\varphi_n = 0
\end{eqnarray}
where $\varphi=\psi_n\exp(-2i\rho^2 t)$. Let us consider the case
of infinite number of lattice sites for which the integrals of
motion in Eq.~(\ref{e1}) are written as
\begin{equation} \label{N}
  N_0 = \sum_{n}(|\varphi_n|^2-\rho^2),
\end{equation}
and
\begin{equation} \label{H}
 H = \sum_{n}\left((\bar{\varphi}_n\varphi_{n-1}-\rho^2)
      -(|\varphi_n|^2-\rho^2)^2\right).
\end{equation}
Moreover, since the problem has  nonzero boundary conditions we
must have, for consistency with Eq.~(\ref{e1}) \cite{KonTMF}, that
\begin{equation} \label{e2}
  \varphi_n\rightarrow\rho\exp[i\Phi_n(t)] \quad \mbox{ at } \quad
  n \rightarrow \pm \infty,
\end{equation}
with the phase $\Phi_n(t)$ defined as
\begin{eqnarray} \label{e4}
  \Phi_n(t)&=&nt\gamma+\chi(t),\\
  \chi(t)&=&2t\left(\frac{\sin \gamma t}{\gamma t}-1 \right).
\label{e5}
\end{eqnarray}
The following gauge transformation \cite{KCV,KonTMF}
\begin{equation} \label{e6}
  q_n=\varphi_n e^{-i\gamma nt},
\end{equation}
allows to rewrite Eq.~(\ref{e1}) as
\begin{equation} \label{e7}
  i\dot{q}_n + e^{i\gamma t} q_{n+1}+ e^{-i\gamma t} q_{n-1} -
  2q_n + 2(\rho^2-|q_n|^2) q_n = 0,
\end{equation}
with boundary conditions
\begin{equation} \label{e8}
  q_n \rightarrow \rho \exp[i\chi(t)], \quad \mbox{at} \quad
  n \rightarrow \pm \infty.
\end{equation}
In the next subsection we use this equation to study the stability
of the background of a BEC array in presence of linear potentials
($\gamma \neq 0$).

\subsection{Linear stability analysis of the background}
\noindent
Let us consider a solution of Eq.~(\ref{e7}) of the form
\begin{equation} \label{e9}
  q_n=[\rho+\alpha_n(t)]\exp[i\chi(t)],
\end{equation}
where $|\alpha_n(t)| \ll \rho$. Linearizing Eq.~(\ref{e7}) around
$\rho\exp[i\chi(t)]$ and expanding in Fourier modes
\begin{eqnarray}
  \beta(k,t) &=& \sum_{n=-\infty}^{\infty}\alpha_n(t)e^{ikn},
  \nonumber \\
  \alpha_n(t)&=& \frac{1}{2\pi}\int_{0}^{2\pi}\beta(k,t)e^{-ikn}dk,
  \nonumber
\end{eqnarray}
we obtain
\begin{equation} \label{e11}
  \frac{d}{d t}|k,t\rangle={\bf T}_k(t)|k,t\rangle,
\end{equation}
where ${\bf T}_k(t)$ is a $4\times 4$ matrix whose nonzero
elements are
\begin{eqnarray}
  T_k^{(12)}&=&-T_k^{(21)}=2[\cos(\gamma t-k)-\cos(\gamma t)-\rho^2],
  \nonumber \\
  T_k^{(34)}&=&-T_k^{(43)}=2[\cos(\gamma t+k)-\cos(\gamma t)-\rho^2],
  \nonumber \\
  T_k^{(14)}&=&T_k^{(23)}=T_k^{(32)}=T_k^{(41)}=2\rho^2, \nonumber
\end{eqnarray}
and with $|k,t\rangle$ a Bloch-Floquet state written in vector form
\begin{eqnarray} \label{e10}
  |k,t\rangle = \left(
  \begin{array}{c}
  \beta_1(k,t) \\
  \beta_2(k,t) \\
  \beta_1(-k,t) \\
  \beta_2(-k,t) \end{array}
  \right)
\end{eqnarray}
(here $\beta(k,t)=\beta_1(k,t) +i \beta_2(k,t)$ with
$\beta_j(k,t)$ real).
\begin{figure}[h]
\vspace{0.5 true cm} \caption{Floquet stability analysis of a
k-modulated background as a function of $\gamma$.  Regions of
stability are designated with "$+$".} \label{figure2}
\end{figure}
\noindent
Since ${\bf T}_k$ is periodic we can use Floquet theory
\cite{Nayfeh} to study the stability. To this end we introduce the
matrix ${\bf S}_k(t)$ as a solution of the equation
\begin{equation} \label{Sk}
  d{\bf S}_k/dt={\bf T}_k{\bf S}_k ,
\end{equation}
satisfying the initial condition ${\bf S}_k(0)={\bf I}$ (here
${\bf I}$ is the $4\times 4$ unit matrix). It then follows  that
the background is unstable  if $|\lambda_j(k)|>1$, where
$\lambda_j(k)$ is the eigenvalue of ${\bf S}_k(2\pi/\gamma)$ (note
that ${\bf S}_k(t)$ is unimodular,  $\det {\bf S}_k(t)\equiv 1$,
so that $\lambda_1 \lambda_2 \lambda_3 \lambda_4=1$). In Fig.
\ref{figure2} the results of Floquet analysis, as obtained from
numerical integrations of Eq.~(\ref{Sk}), are reported (in the
numerical scheme we used a DOPRI8 procedure to integrate
Eq.~(\ref{Sk}) and computed the spectrum of the matrix ${\bf
S}_k$ by reducing it to Hessenberg form \cite{NumRecipes}).
The stability of the background  was also investigated by direct
numerical integrations of Eq.~(\ref{e7}) taking as initial
conditions a uniform background of amplitude $\rho=1$ modulated
by a sine wave of wavenumber $k_n=2 \pi n/L$ (we used  $L=400$
lattice sites).
\begin{figure}[h]
\caption{Modulational stability of the background at $\gamma=0$,
for an initial modulation with  $k\sim 0.31$ }
\label{figure3}
\end{figure}
\begin{figure}[h]
\caption{Modulational instability at $\gamma = 0.5$, for an
initial modulation with  $k\sim 0.31$ }
\label{figure4}
\end{figure}
These results are reported in Figs. \ref{figure3}, \ref{figure4}
for the cases $\gamma=0$, and $\gamma = 0.5$, respectively. We see
that in presence of the linear potential a modulational
instability can develop, in agreement with the Floquet theory
analysis of Fig. \ref{figure2}. A good agreement was found also
for other values of $\gamma$ and other initial conditions (at
$\gamma=0$ the stability was checked only by numerical
integrations of the original equation, since in this case $t =
2\pi/\gamma \rightarrow \infty$ and the Floquet theory becomes
unapplicable). In particular we checked that the modulational
instability of the background  develops at later values of $t$, as
$\gamma$ increases, and that an initial background with $k=\pi$
becomes again stable for $\gamma \sim 5.4$, in agreement with
Fig.\ref{figure2}.

\subsection{Dynamics of small amplitude pulses}

The fact that Eq.~(\ref{e1}) possesses an integral of motion of
type  (\ref{N}), implies that it cannot support solitary waves
\cite{SMK}, i.e. it can not have solutions moving with constant
velocity $v$ and depending on $n, t,$ through the combination
$n-vt$  (this is true also for $\gamma=0$). On the other hand, one
can show that localized smooth excitations of small amplitude can
propagate along the array with a relatively weak distortion. To
investigate this dynamics, it is convenient to introduce a small
parameter $\mu\ll 1$ defined as the square root of the ratio
between the deviation from the background and the background
itself, and consider the case of very small $\gamma$, i.e. we take
$\gamma=o(\mu^3)$. This allow us to look for solutions of the form
\begin{equation} \label{e20}
  q_n=[\rho+\mu^2a]e^{i[\chi+\mu\phi]},
\end{equation}
where $a=a_0+\mu^2a_1+O(\mu^4)$ and
$\phi=\phi_0+\mu^2\phi_1+O(\mu^4)$ are real functions of the {\em
slow} variables $\xi=\mu n$, $T=\mu t$, $\tau=\mu^3 t$, considered
to be continuous and independent.  We can then perform a multiple
scale expansion \cite{KS1} of Eq.~(\ref{e7}), substituting $\gamma
t$ with $\delta\tau$, with $\delta\sim o(1)$. A straightforward
algebra shows that the expansion equation at the first order in
$\mu$ is identically satisfied by the substitution (\ref{e20}).
\begin{figure}[h]
\caption{Time dependence of coefficients of Eq.~(\ref{KdV}) for
$\delta=0.1, \ \rho=1.$ }
\label{figure5}
\end{figure}
The equations at the orders $O(\mu^2)$ and $O(\mu^3)$ are
\begin{equation} \label{e21}
  a_0=-\frac{1}{4 \rho}
  [\partial_T\phi_0+2\sin(\delta\tau)\partial_\xi\phi_0],
\end{equation}
and
\begin{eqnarray} \label{e22}
  \partial_T^2\phi_0+4\sin(\delta\tau)\partial_T\partial_\xi\phi_0 &+&
  \nonumber \\
  4[(\sin(\delta\tau))^2 &-& \rho^2\cos(\delta\tau)]
  \partial_\xi^2\phi_0 =
  0,
\end{eqnarray}
respectively.  It is worth to note that Eq.~(\ref{e22}) at times
$\delta\tau_0=\pm\pi/2$ changes from  hyperbolic (supporting  wave
propagation if $\cos(\delta\tau)>0$) to elliptic. Let us  consider
the dynamics of the small amplitude excitation during the time in
which Eq.~(\ref{e22}) is of the hyperbolic type. We introduce a
new running variable $\zeta=\xi-V(\tau)T$ where
\begin{equation} \label{e25}
  V(\tau)=2\sin(\delta\tau)+2\rho\sqrt{\cos(\delta\tau)}
\end{equation}
can be interpreted as slowly varying velocity of the wave packet
(for the sake of definiteness we have chosen only one branch of
the solution, the other branch being characterized by the
"velocity" with opposite sign). One can then show that
Eq.~(\ref{e22}) is satisfied for arbitrary pair of functions
$a_0(\zeta,\tau)$, $\phi_0(\zeta,\tau)$ linked by
\begin{equation} \label{link}
  a_0(\zeta,\tau)=\frac 12 \sqrt{\cos(\delta
  \tau)}\partial_{\zeta}\phi_0(\zeta,\tau).
\end{equation}
In order to find the dependence of those functions on $\tau$, one
has to consider the equations of the forth and fifth orders in
$\mu$ (more precisely the condition of their compatibility). After
tedious but straightforward calculations, one arrives at the
following Korteweg - de Vries (KdV) equation
\begin{equation} \label{KdV}
  \partial_{\tau} a_0+b_1(\tau)a_0 \partial_{\zeta} a_0
  +b_2(\tau)\partial_\zeta^3 a_0+b_0(\tau)a_0=0,
\end{equation}
with slowly (if $\delta=o(1)$) varying coefficients $b_{j}(\tau)$
given by
\begin{eqnarray} \label{b0b1b2}
  b_0(\tau) &=& \frac 14 \delta\tan(\delta\tau), \nonumber \quad
  b_1(\tau) = 2\frac{3[\cos(\delta\tau)]^{3/2}-
      \rho\sin(\delta\tau)}{\cos(\delta\tau)}, \nonumber \\
  b_2(\tau) &=& \frac{1}{12\rho} \left(4\rho\sin(\delta\tau)+\rho^2
     [\cos(\delta\tau)]^{1/2}-3[\cos(\delta\tau)]^{3/2}\right).
\nonumber
\end{eqnarray}
\begin{figure}[h]
\caption{Evolution of the localized excitation in a BEC array,
governed by Eq.~(\ref{KdV}). The initial wave profile corresponds
to the pure KdV soliton (Eq.~(\ref{pureKdV})) with parameters $u_0
= 1, \ \beta=1.$ } \label{figure6}
\end{figure}
In Fig.\ref{figure6} we report the time evolution of a numerical
solution of Eq.~(\ref{KdV}) with initial condition
\begin{eqnarray} \label{pureKdV}
a_0(\zeta,\tau) &=& 3v {\rm sech}^2(\frac{\zeta-v
\tau}{l_c}),\qquad l_c=\left(4 \frac{\beta}{v}\right)^{1/2}.
\nonumber
\end{eqnarray}
Two important features can be seen from this figure. The first is
the existence of a breaking time $\tau_{br}$  below which the
profile of the wave is distorted but its evolution is smooth (in
Fig.~\ref{figure6} $\tau_{br}\sim 0.5$ ). At $t=\tau_{br}$ the
breaking of the wavefront occurs, after which a train of solitary
pulses (each of them being a KdV soliton) is emitted. This is
similar to the generation of shock waves in a fluid (nonlinear
Schr\"odinger {\em lattice shock waves} were first studied in
Ref.~\cite{KS1}). Note that at the breaking time the coefficient
$b_2(\tau_{br})$ is exactly zero (see Fig.~\ref{figure5}), and
Eq.~(\ref{KdV}) reduces to a known shock dynamics equation
\cite{Witham}. An essential difference between the case at hand
and the case $\gamma=0$ treated in Ref.~\cite{KS1}, however, is
that in this last shock waves can appear only for a given carrier
wave background, while here the linear potential (gravitational
field of BEC arrays) makes them possible to exist for any
wavenumber. Note that for the observation of shock waves it is
important that $\delta$ is small compared with the amplitude of
the background, this ensuring the time at which the group velocity
dispersion becomes negligible, be long enough for shock waves to
develop.

The second important feature  which emerges  from
Fig.~\ref{figure6} is the possibility to have bright pulses
propagating on top of a nonzero background (i.e. on top of a
constant atomic population along the array). These pulses actually
are ES, similar to those predicted in \cite{CKV} for $\gamma=0$.
For $\gamma \neq 0$, these excitations can undergo Bloch
oscillations. To show this we have performed numerical
integrations of Eq.~(\ref{e7}), with a sech initial profile of
large amplitude (i.e. comparable with the background level). The
results are reported in Fig.\ref{bl_osc} (the numerical scheme
used is the same of the previous section). We note that for the
same parameter values, the bright excitation (i.e. above
background) breaks down while the dark one (below background)
remains stable for long times.
\begin{figure}[h]
\caption{Bloch oscillations of bright (a) and dark (b)
localized excitations. Initial conditions: \\
$q_n(0)=\rho \pm 2\eta\,{\rm Sech}[2\eta n] \exp[i\phi_n(0)]$,
$\rho=0.1,$ \ $\eta=0.04$, \ $\gamma=0.1$, \
$\phi_n(0)=-\frac{\pi}{2}n$.}
\label{bl_osc}
\end{figure}
An important conclusion following from these numerical studies is
that bright and dark solitons of KdV type on top of homogeneous
backgrounds, can exist in BEC array with positive scattering
lengths.

\section{Array of BEC with negative scattering lengths}

A distinctive feature of BECs with attractive interactions is the
possibility of collapse when the number of atoms in the condensate
exceeds a critical value $N_c$. From GPE one can predict $N_c
\sim 1400$ for $^7$Li atoms, a value which was confirmed
experimentally in Ref.~\cite{Bradley}. Arrays of optical traps can
be used for manipulation of BECs with negative scattering length,
if their wells are loaded with number of atoms less than this
critical value. In this situation it is reasonable to consider the
dynamics of ES on zero background, as described by
Eq.~(\ref{DNLS1}) with $\sigma=1$. Bloch oscillations of bright ES
were numerically investigated in Ref.~\cite{Scharf}. From these
simulations, a strong dependence of the dynamics on the amplitude
of the initial wave was found (large amplitude wavepackets are
quickly reduced to fragments, while small amplitude excitations
keep their integrity over many oscillation periods). From
numerical studies is also known that the amplitude of a
dynamically localized ES can oscillate in time. Using the method
of the previous section, we can develop analytical considerations
on Bloch oscillations in BEC arrays with negative scattering lengths which
explain the origin of these oscillations. We consider the case of
small amplitudes and small $\gamma$, i.e. we assume, as before,
$\mu=\sqrt{\gamma}$ to be a small parameter of the problem.
Introducing the variable
\begin{equation} \label{newq}
  q_n=\psi_n\exp(-i\gamma n t),
\end{equation}
Eq.~(\ref{DNLS1}) can be written as
\begin{eqnarray} \label{eqq}
  i\dot{q}_n+\cos (\gamma t)(q_{n+1}+q_{n-1}-2q_n)+ \nonumber \\
  i\sin (\gamma t)(q_{n+1}-q_{n-1})+2|q_n|^2 q_n=0,
\end{eqnarray}
whose solution can be searched of the form
\begin{equation} \label{expan}
  q_n=\mu Q(\xi,\tau),
\end{equation}
with $\xi (\tau)=\mu n-x(\tau)$, $\tau=\mu^2t$, and
\begin{equation} \label{eqx}
  x=\frac{2}{\mu}(1-\cos \tau).
\end{equation}
Note that $x(\tau)=O(\mu^3)=O(\gamma^{3/2})$ and thus the
substitution is self-consistent (at $\mu^3t\ll 1$ the term
$\cos(\tau)$ can be expanded in Taylor series). After calculations
analogues to those of the previous section,  one arrives at the
following nonlinear Schr\"{o}dinger (NLS) equation with
periodically varying dispersion
\begin{equation} \label{cdm}
  i\partial_{\tau} Q+\cos(\tau)\partial_{\xi}^{2}Q+2|Q|^2Q=0.
\end{equation}
The dynamics of this equation was  numerically investigated
starting with initial conditions of the form
\begin{equation} \label{Q0}
  Q(\xi,0)=2\eta{\rm sech}(2\eta(\xi-\xi_0)).
\end{equation}
For the numerical code we used a split-step fast Fourier transform
technique \cite{Agrawal} with a time step $\Delta \tau=0.001$ and
a space step $\Delta \xi=0.02$ (in normalized units),
corresponding to 1024 grid points in the discretization domain
taken as -10 $\div$ 10 (absorbing boundary conditions were used to
simulate an infinite domain). The accuracy of the scheme was
checked by monitoring  the conservation of the number of atoms and
of the Hamiltonian,  which was within  $\pm$ 0.1 \% for all runs.
The results are depicted in Fig.\ref{figure8}.
\begin{figure}[htb]
\caption{(a) Evolution of the localized excitation governed by
Eq.~(\ref{cdm}), for $\eta=0.5.$ and (b) Decay of its amplitude.}
\label{figure8}
\end{figure}
From Fig.~\ref{figure8}a) we see that while the ES is executing
spatial oscillations, its amplitude is oscillating in time. In
Fig.~\ref{figure8}b) we report the time dependence of the center
of the wave on a longer time scale, from which we see that the
amplitude while oscillating is also decaying (note, however, that
this dynamics reproduces the dynamics  of a  real array only for
times $\tau$ such that $\Delta \tau <1$). A more detailed
numerical investigation shows that the amplitude can be either
decaying or growing in time, depending on initial conditions (the
growth can reach also the $60$\% of the initial amplitude of the
pulse). This implies that if at $t=0$ the number of atoms in one
well is below the critical value $N_0(t=0)< N_{cr}$ for collapse,
the growth in amplitude due to the linear potential, can induce
collapse after a time $t=t_{cr}$ at which $N_0(t_{cr} )=N_c$.

\section{Conclusions}
The dynamical properties of BEC arrays with positive and negative
scattering lengths have been studied in the framework of a
discrete nonlinear Schr\"odinger equation derived from the mean
field GPE with a tight-binding approximation. The interplay
between macroscopic inter site tunnelling and nonlinear
self-trapping was shown to be responsible for the appearance of
different types of DMS. In particular we showed, both analytically
and numerically, that for a small ratio between the tunnel
coupling constant and the nonlinear interatomic interactions, ILM
solutions can exist. For BEC arrays with non-zero backgrounds the
modulational stability problem was investigated and the existence
of bright and dark ES was discussed. The problem of Bloch
oscillations of envelope solitons in arrays with positive
scattering lengths was also analytically and numerically
investigated. We showed that at the lower orders of a perturbative
expansion, the dynamics of small amplitude excitation evolves
according to a KdV equation with time dependent coefficients. In
this case the possibility of shock waves formation, in BEC arrays
with positive scattering lengths, was explicitly displayed. In the
case of arrays with negative scattering lengths, the dynamics of
small amplitude excitations was described in terms of a nonlinear
Schr\"odinger equation with a periodically varying dispersion.
This equation was used to show the presence of amplitude
oscillations during Bloch oscillations, as well as decay or growth
of the excitations, depending on the initial conditions. The
results of this paper clearly show the complexity and the wide
range of behaviors which can arise in the system.

In closing this paper we feel compelled to discuss to which extend
the phenomena presented in this paper could be observable in real
BEC arrays. Our model is based on a tight binding approximation,
this putting restrictions on the shape of the wavefunctions, as
well as, on the potential profile, to be used. Although a detailed
analysis of the experimental settings for which  the tight binding
approximation is accurate has not yet been done, there are
experimental situations in which this is obviously true. Thus, for
example, if the potential barrier is wide and tall enough to
reduce the tunnelling probability among adjacent BECs, the
overlapping of the wavefunctions is certainly  small and the model
is accurate. In this case DMS of the type discussed above should
appear.

In particular, the possibility to observe ILMs in real experiments
should not be overlooked. We remark that from
Eqs.~(\ref{normalization2}), (\ref{renorm}), it follows that the
peak amplitude, i.e. $\max(\phi_n(t))$, is of order $O(1)$, so
that the atomic population can be localized on very few sites of
the array. In the experiment in Ref.~\cite{Kasevich}, the peak
density was $n_0 = 10^{13}/$cm$^3$, the mean field energy $g_0 n_0
\approx k_B \cdot 4$nK ($k_B$ being the Boltzmann constant), this
giving the parameter $\kappa$ in Eq.~(\ref{klim}) to be
$\kappa\approx 1$. To observe ILM one needs to have at least
$\kappa \sim 0.1$. This can be achieved either by increasing the
number of atoms in each well (from $10^3$ to $10^4$), or by
reducing the tunnelling constant $K$, which exponentially depends on
the lattice constant (distance between potential wells). When the
number of atoms in the wells is increased, however, a loss of
coherence of the condensate can occur, a phenomenon observed in
the experiments reported in Ref.~\cite{Mewes}. On the other hand,
$\kappa$ can be reduced also by changing the atomic scattering
length $a_s$ using the Feshbach resonances \cite{Ino}, so that it
should be possible to find experimental settings for which ILM can
be observed.

As to ES on finite backgrounds  discussed in Section IV, we remark
that the number of atoms involved in these KdV solitons can be
estimated as $N_s = 2\mu\rho \int_{-\infty}^{\infty}
a(\zeta/l_c)d\zeta = 4\mu u_0 \rho l_c .$ For the parameters used
in the numerical simulations of Fig.~\ref{bl_osc}, with $\delta
=\mu^3 = 0.1, u_0 = \beta = 1$ we have $\mu \approx 0.46, l_c
=3.464$, and for an array of Rb atoms with 1000 atoms in each well,
we have $N_s \sim 6370$ atoms. These bright (dark) solitons on top
of a background, could be created by using a laser beam applied
for short time to a uniform BEC array, so to create a local
enhancement (depletion) of the potential. In this case the
dependence of the soliton velocity on its amplitude is an
interesting parameter to measure since, in contrast with nonlinear
Schr\"odinger solitons, the velocity of KdV solitons depends on
the amplitude of the wave (number of the atoms in the condensate).
This could provide a way to experimentally check our results.

We hope that experiments in this direction will be soon performed.

\section*{Acknowledgments}

F.Kh.A., B.B.B. and S.A.D. are grateful to the US CRDF (Award ZM2-2095)
for partial financial support. V.V.K. acknowledges support from
FEDER and Program PRAXIS XXI, grant N$^o$ Praxis/P/Fis/10279/1998
and the EC grant  HPRN-CT-2000-00158. F.Kh.A. and V.V.K. acknowledge
NATO fellowship program for the support of cooperative work. M.S.
thanks the "Centro de F\'{\i}sica da Mat\'eria Condensada",
University of Lisbon, for a one month Visiting Professorship, the
EC grant HPRN-CT-1999-00163, and the MURST-PRIN-2000 Initiative,
for financial support.


\begin{references}
\bibitem{commento1}In some literature the word discrete breather
appears as synonymous of ILM. Since breathers solutions close to
the continuum (i.e. localized on many lattice sites) can also
exist in these systems and thus can be interpreted as envelope solitons,
we prefer not to introduce this terminology.
\bibitem{commento2}One can estimate from Fig. 4 of
Ref.~\cite{trombettoni} a width
of their excitations of about $30$ lattice sites.
\end{references}
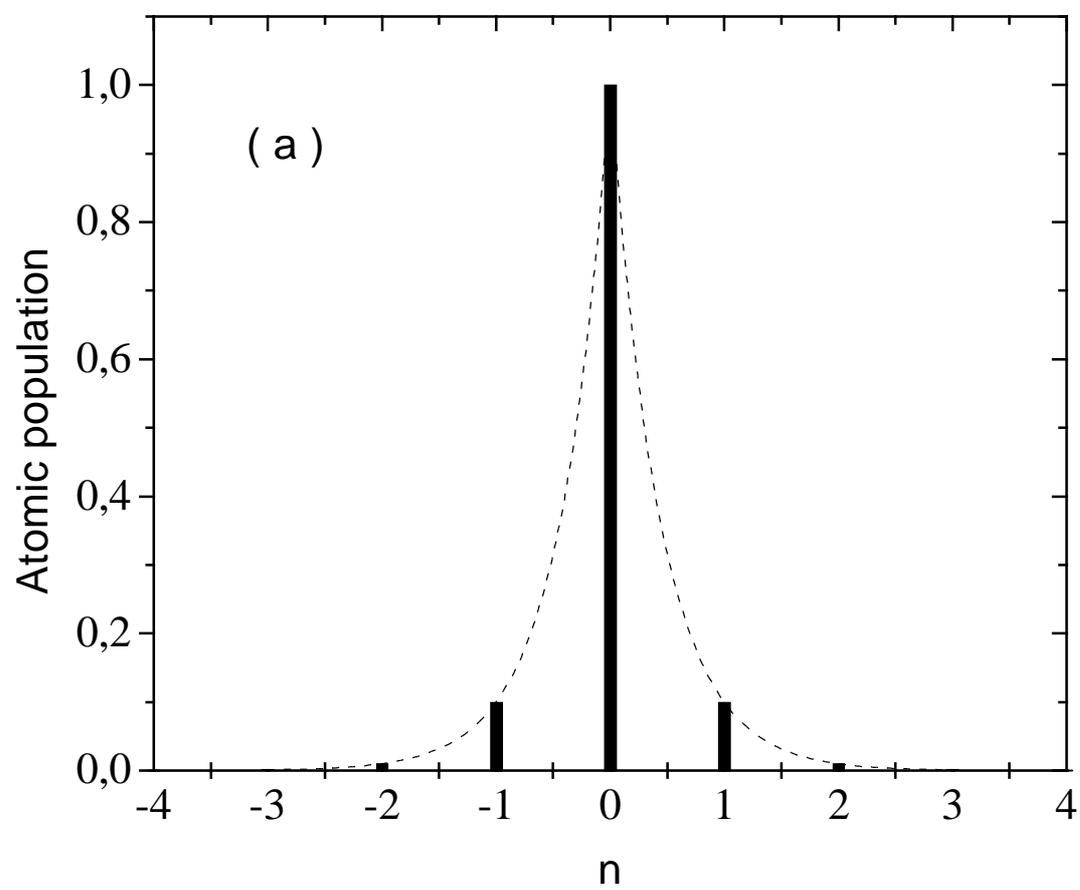
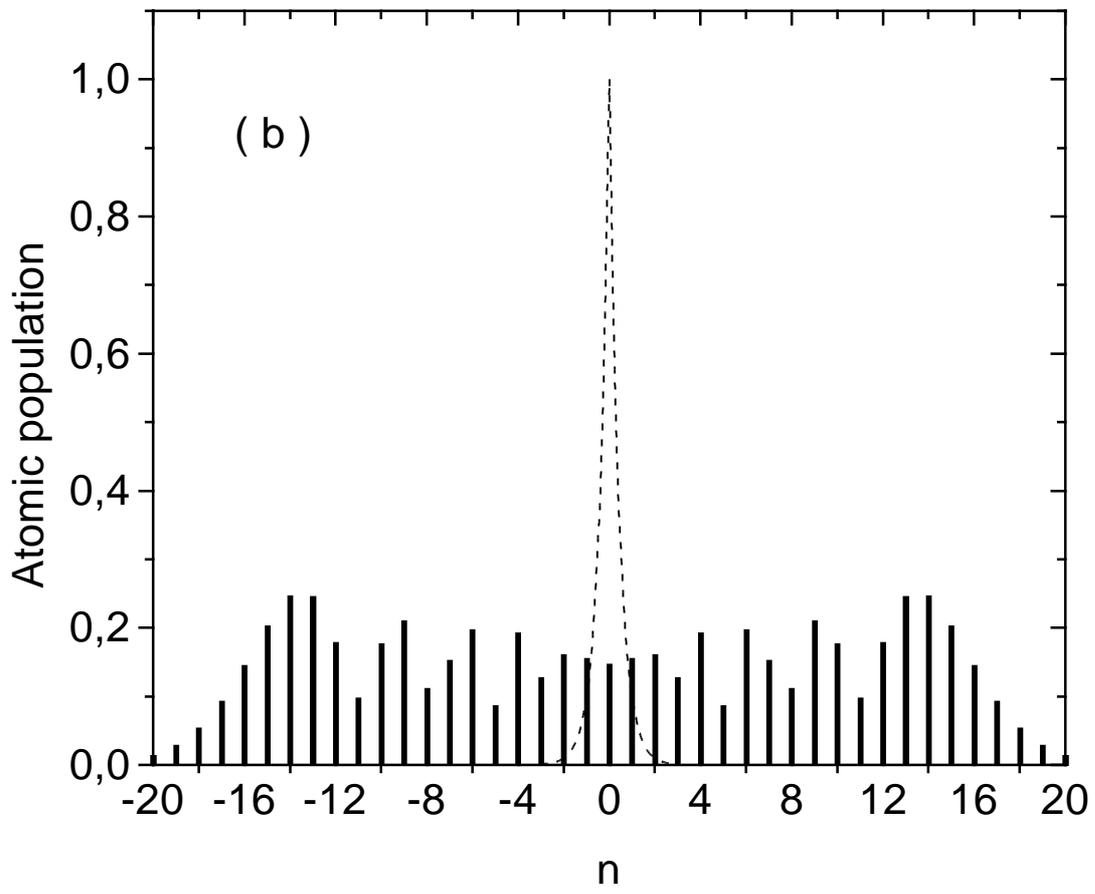
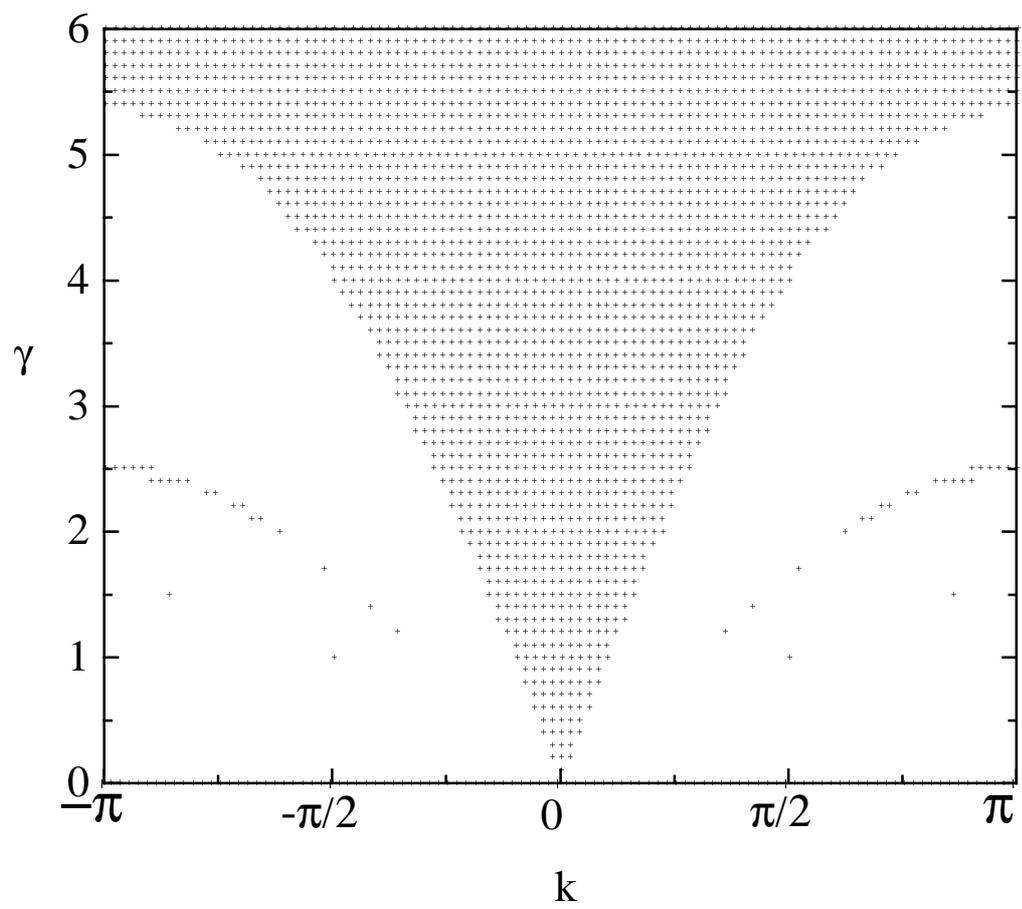
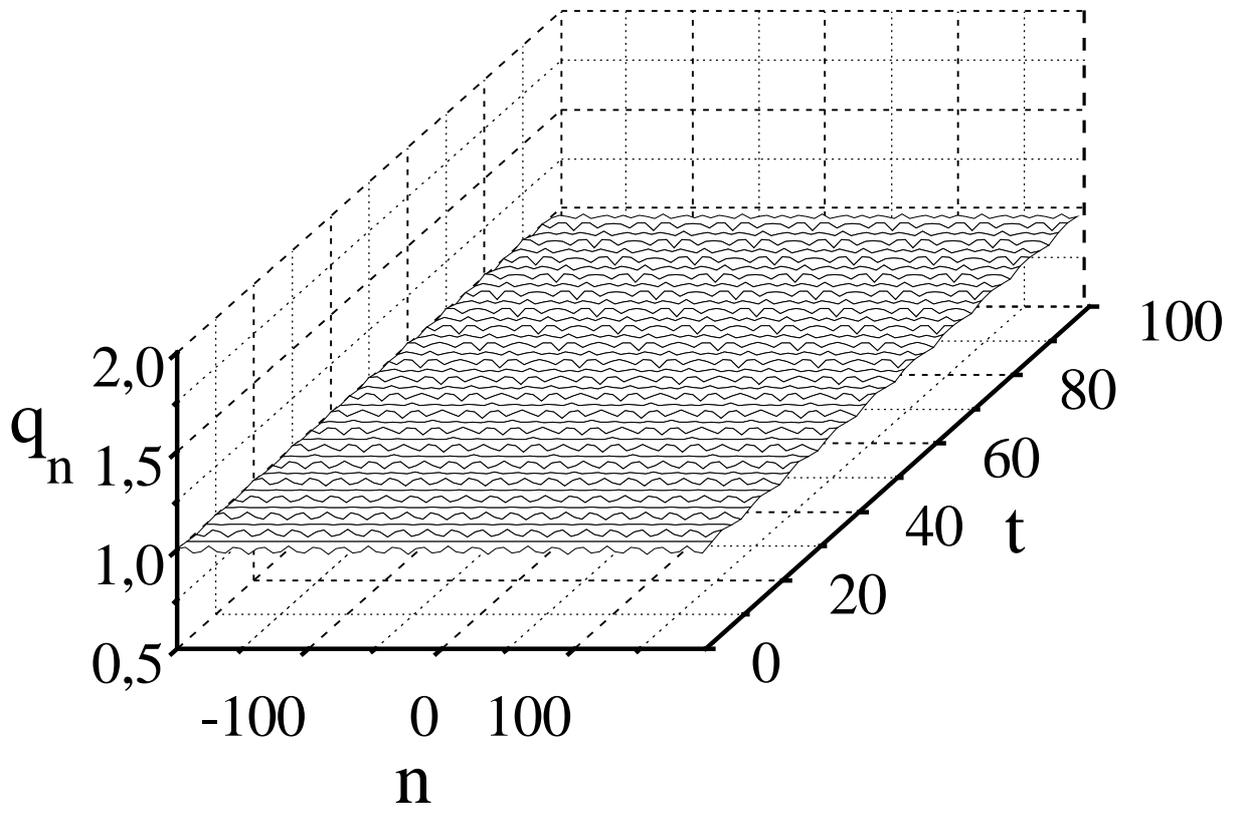
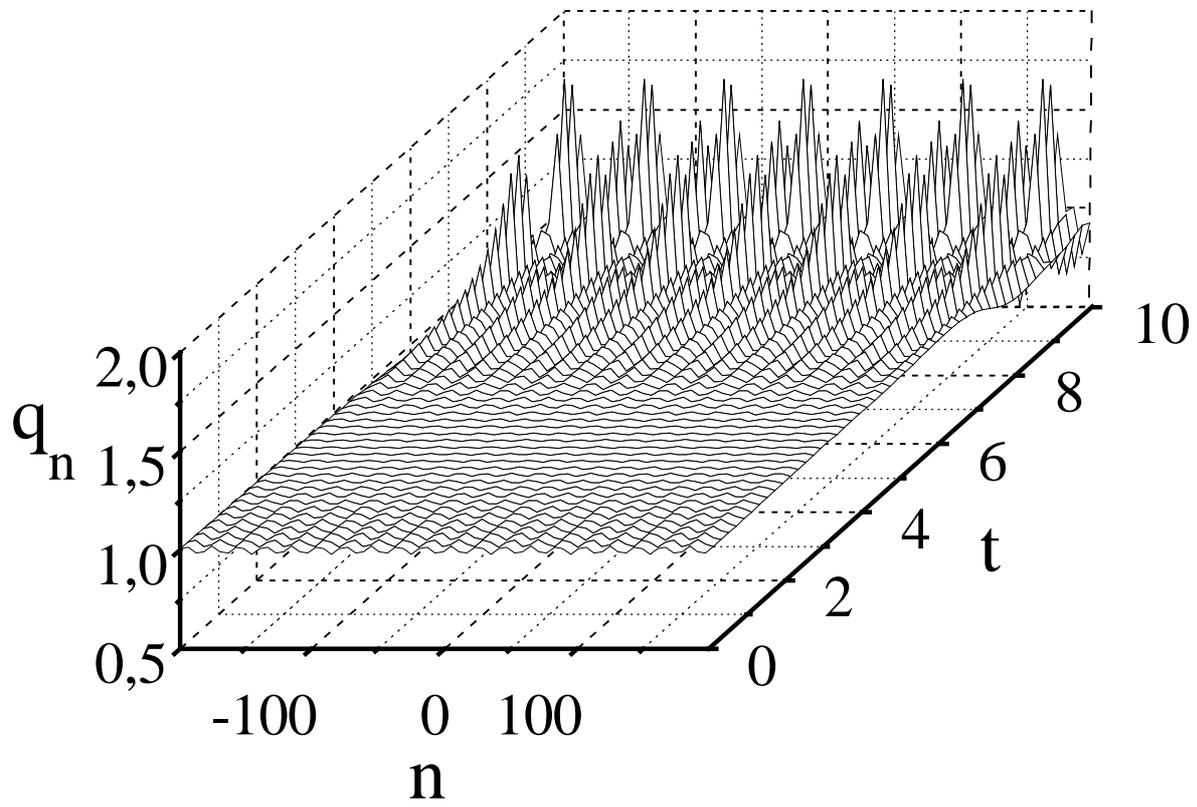
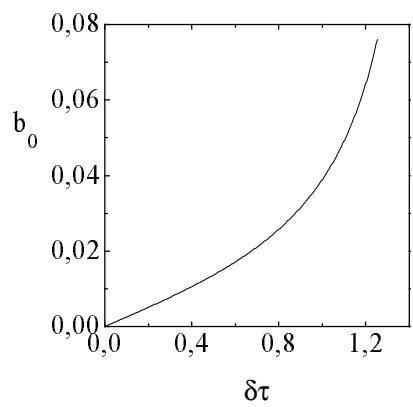
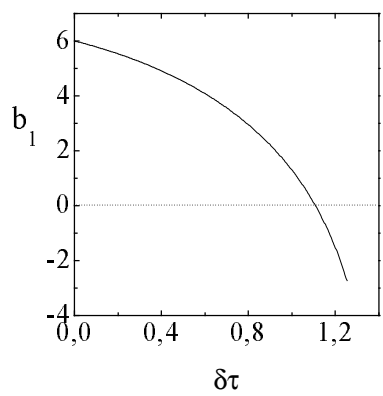
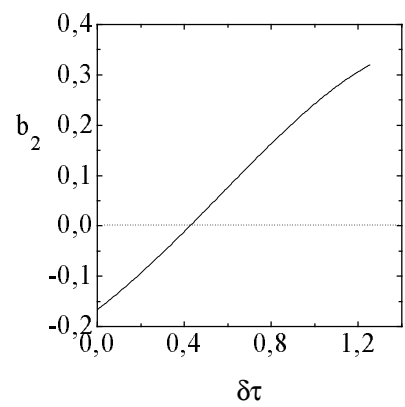
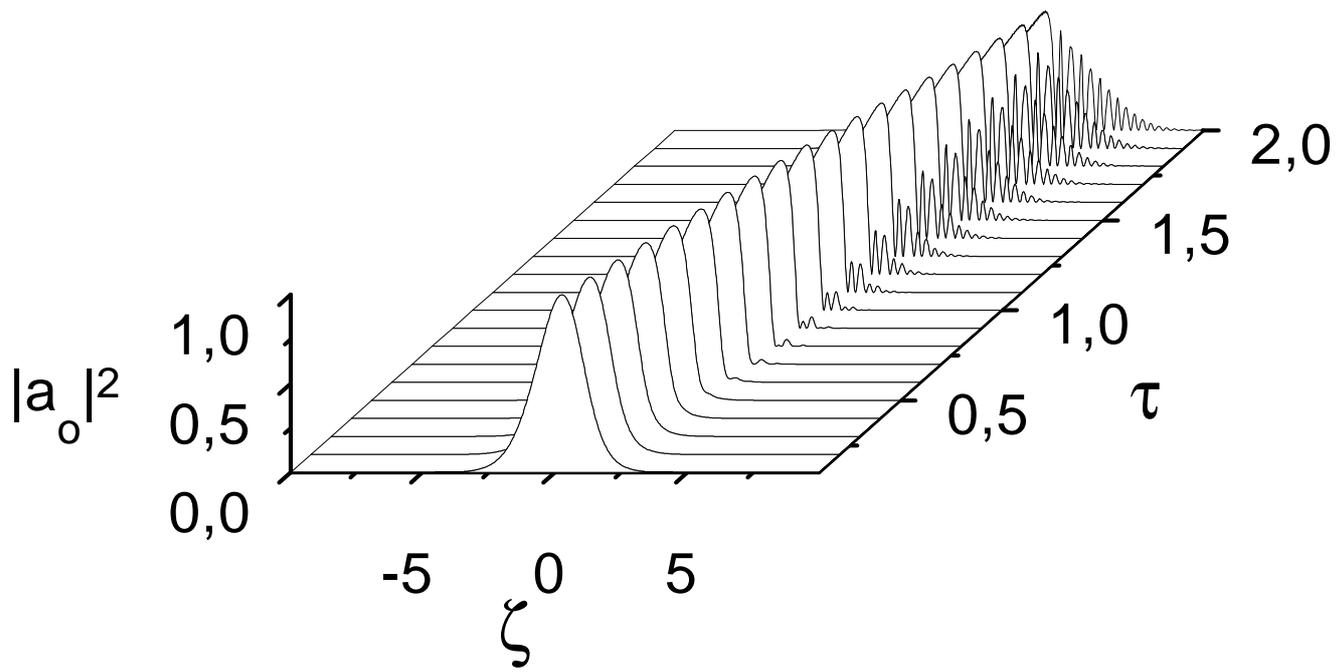
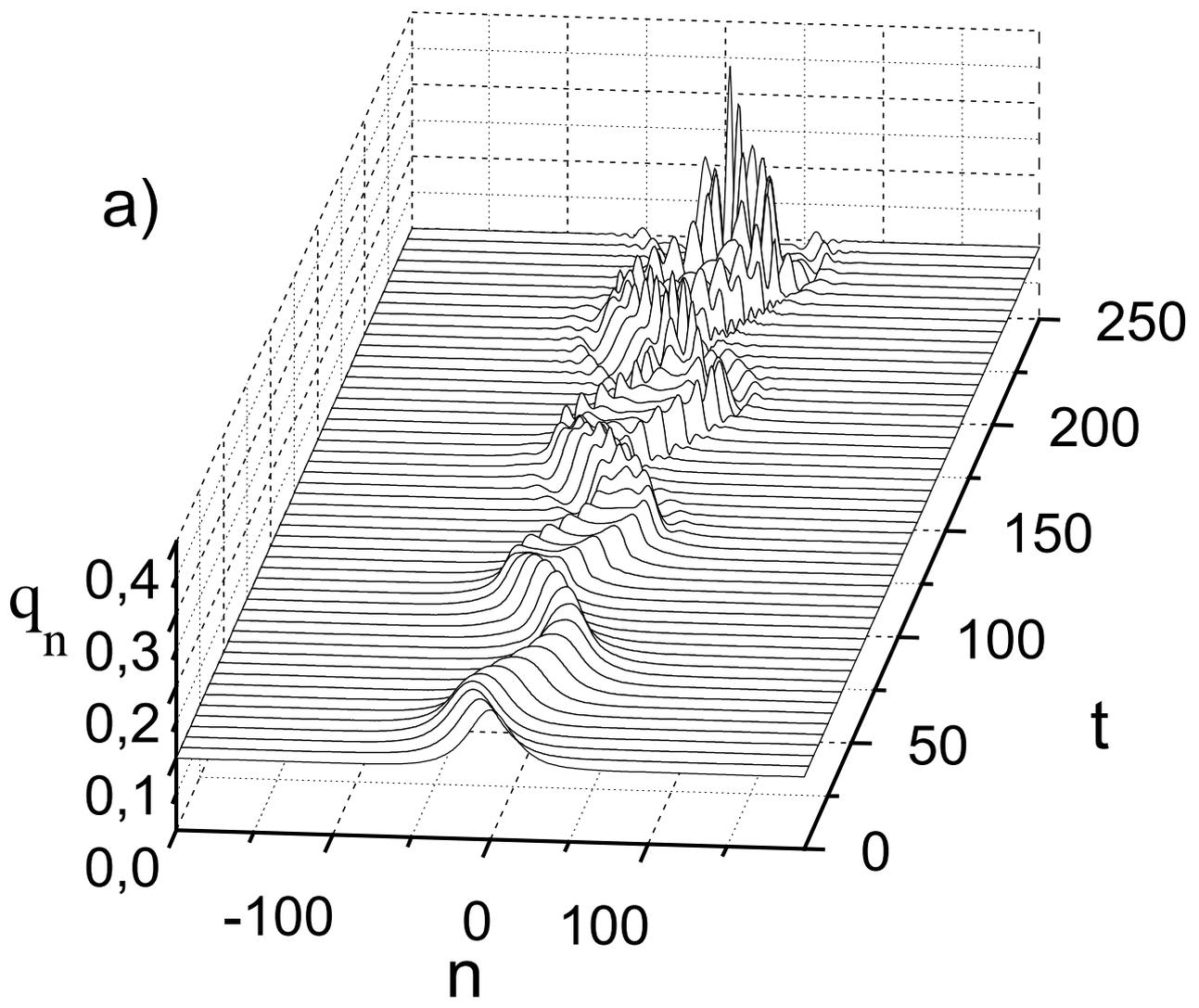
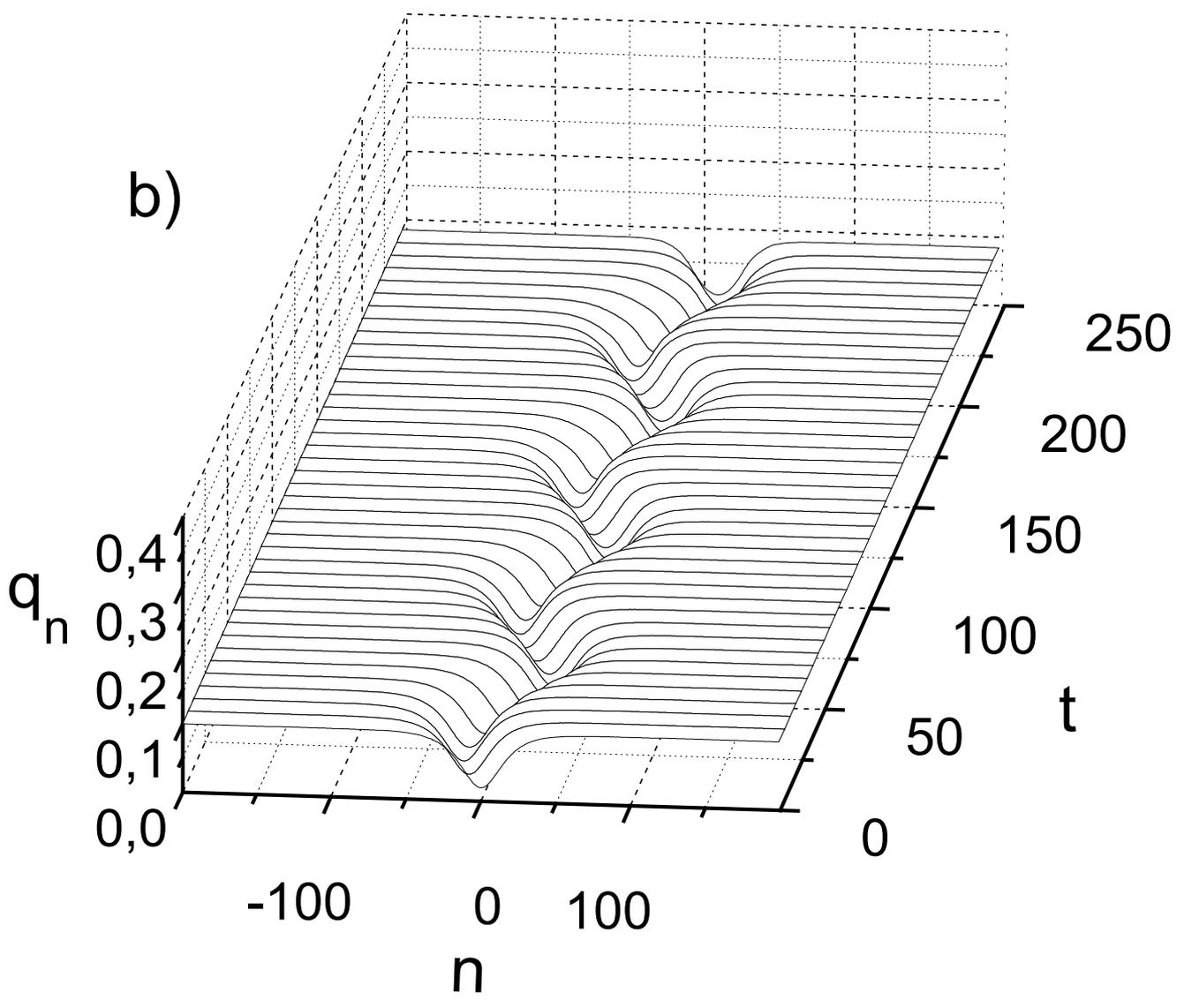
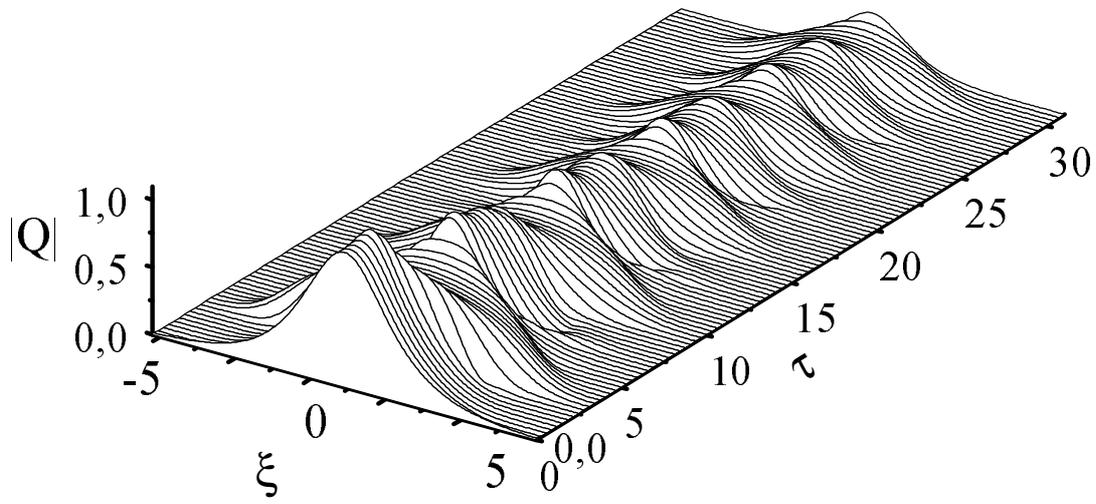
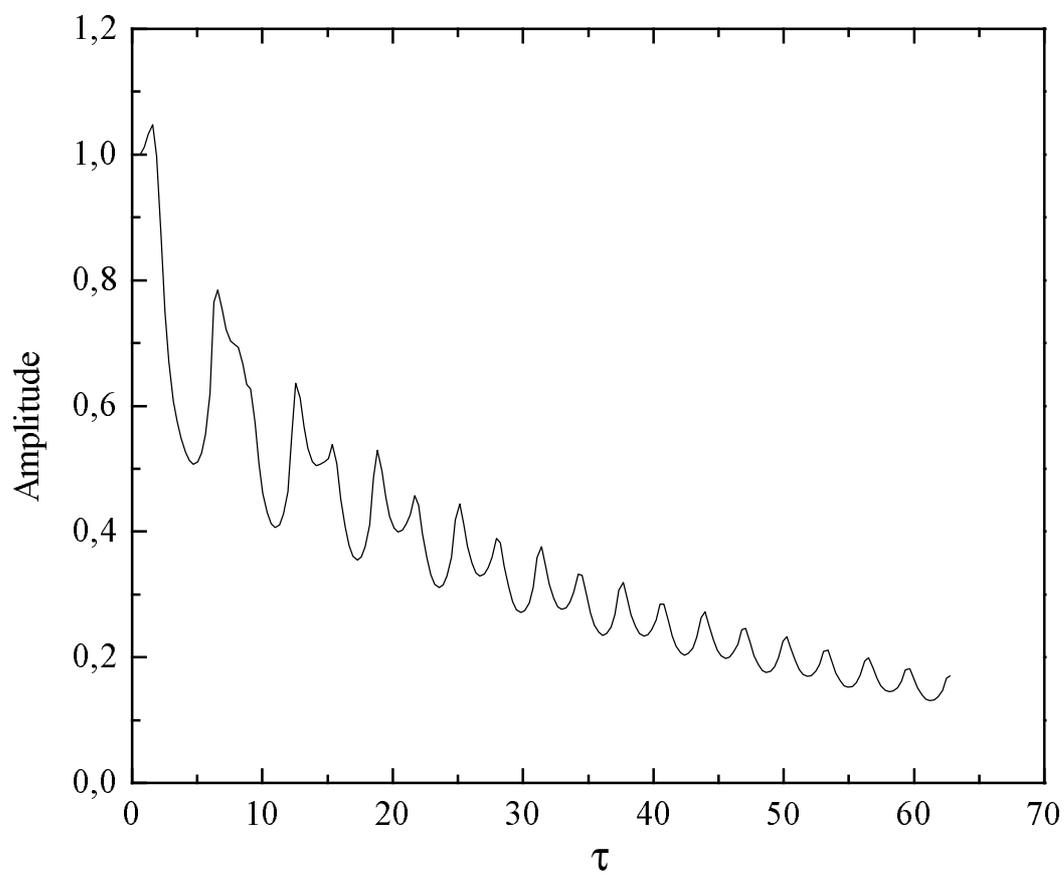
\end{document}